\documentclass[aps,prb,reprint,superscriptaddress,floatfix,showpacs]{revtex4-1}
\usepackage[svgnames,usenames]{xcolor}
\usepackage{amssymb}
\usepackage{amsmath}
\usepackage{textcomp}
\usepackage{graphicx}
\usepackage{multirow}
\usepackage{float}
\usepackage{epstopdf}
\usepackage{natbib}
\usepackage{url}
\usepackage[normalem]{ulem}
\usepackage[bookmarksnumbered,pdfpagelabels=true,plainpages=false,colorlinks=true,linkcolor=blue,citecolor=blue,urlcolor=blue]{hyperref}
\newcommand{\del}{\textcolor{red}\bgroup\markoverwith{\textcolor{red}{\rule[+0.5ex]{2pt}{0.4pt}}}\ULon}
\newcommand{\todol}{\bgroup\markoverwith{\textcolor{red}{\rule[-0.5ex]{2pt}{0.4pt}}}\ULon}

\begin{document}

\title{Spin-Wave Modes in Transition from a Thin Film to a Full Magnonic Crystal} 

\author{M.\ Langer}
\affiliation{Helmholtz-Zentrum Dresden -- Rossendorf, Institute of Ion Beam Physics and Materials Research, Bautzner Landstr.\ 400, 01328 Dresden, Germany}
\affiliation{Institute for Physics of Solids, Technische Universit\"at Dresden, Zellescher Weg 16, 01069 Dresden, Germany}
\affiliation{Swiss Light Source, Paul Scherrer Institute, 5232 Villigen, Switzerland}

\author{R.\ A.\ Gallardo}
\affiliation{Departamento de F\'isica, Universidad T\'ecnica Federico Santa Mar\'ia, Avenida Espa$\tilde{n}$a 1680,
2390123 Valpara\'iso, Chile}
\affiliation{Center for the Development of Nanoscience and Nanotechnology (CEDENNA), 917-0124 Santiago, Chile}

\author{T.\ Schneider}
\affiliation{Helmholtz-Zentrum Dresden -- Rossendorf, Institute of Ion Beam Physics and Materials Research, Bautzner Landstr.\ 400, 01328 Dresden, Germany}
\affiliation{Department of Physics, Technische Universit\"at Chemnitz, Reichenhainer Str.\ 70, 09126 Chemnitz, Germany}

\author{S.\ Stienen}
\affiliation{Helmholtz-Zentrum Dresden -- Rossendorf, Institute of Ion Beam Physics and Materials Research, Bautzner Landstr.\ 400, 01328 Dresden, Germany}

\author{A.\ Rold\'an-Molina}
\affiliation{Universidad de Ays\'en, Calle Obispo Vielmo 62, Coyhaique, Chile}

\author{Y.\ Yuan}
\affiliation{Helmholtz-Zentrum Dresden -- Rossendorf, Institute of Ion Beam Physics and Materials Research, Bautzner Landstr.\ 400, 01328 Dresden, Germany}

\author{K.\ Lenz}
\affiliation{Helmholtz-Zentrum Dresden -- Rossendorf, Institute of Ion Beam Physics and Materials Research, Bautzner Landstr.\ 400, 01328 Dresden, Germany}

\author{J.\ Lindner}
\affiliation{Helmholtz-Zentrum Dresden -- Rossendorf, Institute of Ion Beam Physics and Materials Research, Bautzner Landstr.\ 400, 01328 Dresden, Germany}

\author{P.\ Landeros}
\affiliation{Departamento de F\'isica, Universidad T\'ecnica Federico Santa Mar\'ia, Avenida Espa$\tilde{n}$a 1680,
2390123 Valpara\'iso, Chile}
\affiliation{Center for the Development of Nanoscience and Nanotechnology (CEDENNA), 917-0124 Santiago, Chile}

\author{J.\ Fassbender}
\affiliation{Helmholtz-Zentrum Dresden -- Rossendorf, Institute of Ion Beam Physics and Materials Research, Bautzner Landstr.\ 400, 01328 Dresden, Germany}
\affiliation{Institute for Physics of Solids, Technische Universit\"at Dresden, Zellescher Weg 16, 01069 Dresden, Germany}

\date{\today}

\begin{abstract}
Surface-modulated magnonic crystals are the natural link between continuous films with sinusoidal spin-wave eigenmodes and one-dimensional magnonic crystals composed of individual nanowires. Nevertheless, the transformation process of the spin-wave modes in this transition remains yet unclear. Here, spin-wave modes in their entire transition from a flat film to a `full' (one-dimensional) magnonic crystal are studied by ferromagnetic resonance (FMR) and micromagnetic simulations. For this purpose, the surface of a pre-patterned thin permalloy film was sequentially ion milled resulting in hybrid structures, referred to as surface-modulated magnonic crystals, with increasing modulation depth. After each step, FMR measurements were carried out in backward-volume and Damon-Eshbach geometry. The evolution of each spin-wave resonance is studied together with the corresponding mode profile obtained by micromagnetic simulations. Simple rules describing the transition of the modes from the film to the modes of the full magnonic crystal are provided unraveling the complexity of spin-wave states in these hybrid systems.

\end{abstract}

\pacs{}

\maketitle 

\section{Introduction}
\label{Int}
Periodically patterned magnetic materials with periodicities ranging from micrometers down to several tens of nanometers are referred to as magnonic crystals (MCs).\cite{Vasseur1996,Nikitov2001,Neusser2009,Kruglyak2010,Gubbiotti2010,Lenk2011,Krawczyk2014,Chumak2015,Mansurova2017} In the last decade, this group of meta-materials, such as bi-component systems,\cite{Ma2012,Tacchi2012,Yu2013,Saha2013,Mruczkiewicz2013,Obry2013,Rychly2017} free standing structures,\cite{Topp2010,Ding2011,Duerr2012,Schwarze2013,Huber2013,Montoncello2013} and continuous films with periodic structures on top, also known as surface-modulated magnonic crystals (SMMCs),\cite{Barsukov2011,Landeros2012,Liu2013,Gallardo2014,Kakazei2014,Langer2016,Khanal2017,Heimbach2018} experienced a growing scientific interest. As spin waves (SWs) offer unique properties such as charge-less propagation and high group velocities, there are multiple applications conceivable since industry is in need for higher efficiencies as well as high performances in information technology including the transport and processing of data.\cite{Cuykendall1987,Schneider2008,Lee2008,Khitun2010,Khitun2011,Berut2012,Khitun2012,Sato2013,Jamali2013,Klingler2014,Chumak2015} The possibility to manipulate the band gaps\cite{Kostylev2008,Lee2009,Wang2009,Wang2010,Ma2012,Di2013,Krawczyk2013,Kumar2014} and to tailor the SW properties paves the way for many applications based on magnonic devices,\cite{Inoue2011,Rana2018} such as magnonic filters,\cite{Kim2009} switches\cite{Khitun2010,Vogt2014,Balinskiy2018}, grating couplers\cite{Yu2013} and transistors.\cite{Chumak2014} Moreover, fast developments in spintronics and spin-caloritronics\cite{Bauer2012} hold out the perspective of novel promising hybrid-topics\cite{Chumak2015} in the future, where the unique properties of MCs are combined with functional entities, such as recently shown for spin-torque oscillators.\cite{Urazhdin2014a}


In an ongoing miniaturization process, future devices will require small SW wavelengths in an exchange-dominated regime.\cite{Liu2018} For this reason, a small base periodicity of a few hundred nanometers or less is desired with characteristic SW mode wavelengths of several tens to hundreds of nanometers. There are several advanced methods to investigate the transition from a thin film to a MC with a pronounced modulation, such as e.g.\ thermal landscape modulation\cite{Vogel2015} or periodic Oersted-fields of current-carrying meander structures.\cite{Chumak2009-JPD,Chumak2010,Bai2011,Wang2017} However, these approaches are so far non-applicable to periodicities of 500~nm and below.

\begin{figure}[t]
\includegraphics[width=0.95\columnwidth]{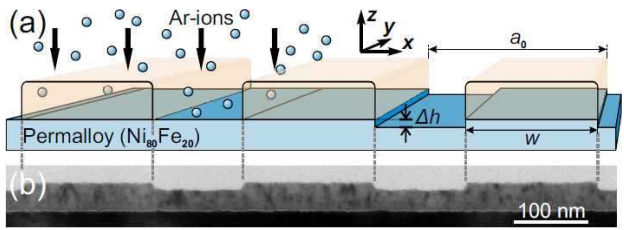}
\caption{\label{experiment} (Color online) (a) Sketch of the sequential fabrication of several SMMCs by ion-milling of a single sample. (b) TEM image of a sample milled to $\Delta d = 10$~nm surface modulation.} 
\end{figure}

This work complements these studies in focusing on transitional systems which offer even richer SW spectra with adjustable amplitudes. The transition of these modes from the film limit to full MC limit is studied by incrementally introducing periodic trenches into the film surface. The resulting structures [sketched in Fig.~\ref{experiment}(a)] can be seen as a periodic array of wires on top of a thin film and are referred to as surface-modulated magnonic crystals.

In table~\ref{tab1}, the literature is summarized in which SMMCs are employed under a systematic variation of the modulation height. This summary reveals the diverging structural properties and the variety of scientific objectives for which these structures were used. As most of the previous investigations cover only a fraction of the full transition (column 4 in table~\ref{tab1}), this work is intended to close the gap between MCs in the surface-perturbation regime\cite{Landeros2012,Gallardo2014} and strongly modulated systems.\cite{Liu2013,Kakazei2014} It is important to furthermore note that most previous work aimed at different objectives, such as the SW transmission of surface-patterned waveguides,\cite{Chumak2008,Chumak2009-JAP,Chumak2009-APL} the optimization of the reconfigurability of the magnonic properties\cite{Liu2013,Kakazei2014}, and the study of SWs in SMMCs in the perturbation regime.\cite{Landeros2012,Gallardo2014} Nonetheless, the work that is closer related to this topic concerns either the numerical analysis of SW resonances together with the respective mode profiles\cite{Aranda2014} or the theoretical and numerical calculations of the locally varying internal demagnetizing fields which act on the SWs opening band gaps and showing flat bands around the backward-volume geometry.\cite{Gallardo2018} This experimental study is meant to complement these theoretical works and shall provide a comprehensible interpretation of the evolution of SW modes during different levels of surface modulation.

Up to now mainly the two limits---the thin film limit\cite{Barsukov2011, Landeros2012, Gallardo2014, Langer2016} and the `full' MC\cite{Demokritov2001,Roussigne2001,Gubbiotti2007,Kostylev2008,Wang2009,Tacchi2010,Wang2010,Saha2015,Belmeguenai2016}---were intensively studied. This work is focused on the dynamic properties of transitional systems---SMMCs. For this purpose, a stripe-patterned thin permalloy (Ni$_{80}$Fe$_{20}$) film was sequentially ion-milled and characterized after each milling step using broadband ferromagnetic resonance (FMR). This allows to study SMMCs with different modulation heights using the same sample. The SW resonances and SW mode profiles are compared to the results of micromagnetic simulations in order to study the transition of SW modes from film modes to the modes in a one-dimensional MC. The huge variety of modes in SMMCs, the strong coupling between them, and the distortion of the mode profiles due to the hybridization and inhomogeneous internal demagnetizing fields hinder a straightforward interpretation. To disentangle these effects, the approach to follow the modes under an incremental increase of the modulation height becomes particularly favorable.

The manuscript is organized as follows. Details about the sample fabrication and the measurement technique are given in Sec.~\ref{Exp} followed by Secs.~\ref{BV} and \ref{DE} containing the results for the backward-volume (BV) and the Damon-Eshbach (DE) geometry, respectively. Section~\ref{Con} summarizes the results.

\section{Experiment}
\label{Exp}

The experiments are based on a polycrystalline $d = 36.8$~nm thin permalloy (Ni$_{80}$Fe$_{20}$) film deposited by electron beam deposition on surface-oxidized Si(001) substrate. The surface of the film was lithographically stripe patterned using ma-N~2401 negative resist with a wire width (resist covered wires) of $w = 140$~nm and a periodicity of $a_0=300$~nm. In order to incrementally remove the magnetic material between the resist-covered stripes, sequential Ar-ion milling was employed. The procedure is schematically depicted in Fig.~\ref{experiment}(a). Figure~\ref{experiment}(b) shows the corresponding cross-section of a patterned permalloy film after milling $\Delta d = 10$~nm into the film. Altogether, the film was milled five times until an array of separate wires, i.e.\ a `full' MC, was achieved. 
\begin{figure}[t]
\includegraphics[width=0.95\columnwidth]{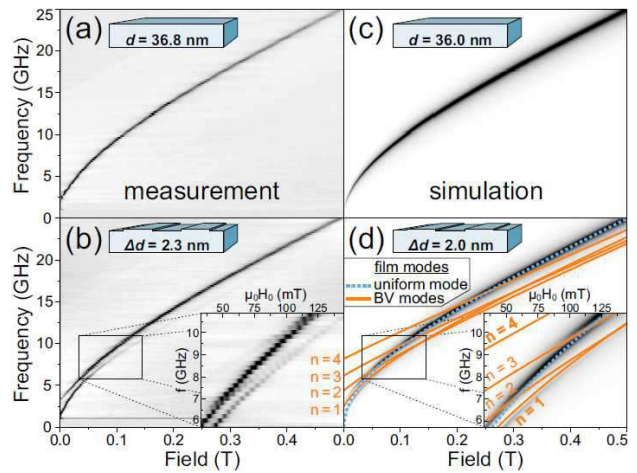}
\caption{\label{FilmLimit} (Color online) Measurement and simulation of the frequency-field dependence of (a),(c) a flat film and (b),(d) a surface-modulated magnonic crystal with tiny modulations ($\Delta d=2$~nm). Orange lines represent the spin waves' frequency-field dependences calculated from Eq.~(\ref{BVdisp}).}
\end{figure}

After each ion-milling step, the frequency-field-dependence $f(H_0)$ was measured using a broadband ferromagnetic resonance (FMR) setup, as described in reference~[\onlinecite{Langer2016}]. 
An FMR pre-characterization of the thin film properties was carried out prior to the sequential ion milling yielding the saturation magnetization $\mu_{0} M_{\mathrm{s}}=0.9236$~T, the $g$-factor $g=2.11$ and the exchange stiffness $D=23.6$~Tnm$^2$.\cite{Langer2016, Langer2017} It shall be noted that the investigations concentrate on symmetric modes only, since the dynamic measurements are carried out using a symmetric excitation scheme.
\begin{table*}[th]
\caption{Overview of previous investigations of systematic modulation $\Delta d/d$ variations on the spin-wave properties of SMMCs. The following abbreviations are used: SWT---spinwave transducer, TMS---two-magnon scattering, PWM---plane-wave method.}
\label{tab1}
\centering
\begin{tabular}{ccccccccccccccc}
\hline\hline
\multirow{2}{*}{authors / year} & ~ & 1D / 2D & ~ & thickness & ~ & modulation & ~ & \multirow{2}{*}{material} & ~ & period & ~ & main & ~ & studied \\
& ~ & SMMC & ~ & $d$ (nm) & ~ & $\Delta d/d$ (\%) & ~ & & ~ & $a_0$ (\textmu m) & ~ & method & ~ & quantity \\
\hline
Chumak \textit{et al.} 2008 [\onlinecite{Chumak2008}] & ~ & 1D & ~ & 5\,500 & ~ & 0--16.4 & ~ & YIG & ~ & 300 & ~ & SWT & ~ & SW transmission\\
\multirow{2}{*}{Chumak \textit{et al.} 2009 [\onlinecite{Chumak2009-JAP}]} & \multirow{2}{*}{~} & \multirow{2}{*}{1D} & \multirow{2}{*}{~} & 5\,500 & \multirow{2}{*}{~} & 0--16.4 & \multirow{2}{*}{~} & \multirow{2}{*}{YIG} & \multirow{2}{*}{~} & \multirow{2}{*}{300} & \multirow{2}{*}{~} & \multirow{2}{*}{SWT} & \multirow{2}{*}{~} & \multirow{2}{*}{SW transmission}\\
& & & & 14\,000 & & 0--12.9 & & & & & & & &\\
Chumak \textit{et al.} 2009 [\onlinecite{Chumak2009-APL}] & ~ & 1D & ~ & 5\,500 & ~ & 0--41.8 & ~ & YIG & ~ & 300 & ~ & SWT & ~ & SW transmission\\
Landeros \textit{et al.} 2012 [\onlinecite{Landeros2012}] & ~ & 1D & ~ & 30 & ~ & 0--6.7 & ~ & Permalloy & ~ & 0.25 & ~ & TMS Theory & ~ & FMR response\\
Liu \textit{et al.} 2013 [\onlinecite{Liu2013}] & ~ & 2D & ~ & 60--120 & ~ & 50--100 & ~ & Permalloy & ~ & 0.62 & ~ & FMR & ~ & FMR response\\ 
Kakazei \textit{et al.} 2014 [\onlinecite{Kakazei2014}] & ~ & 1D & ~ & 30--60 & ~ & 50--100 & ~ & Permalloy & ~ & 0.4 & ~ & FMR & ~ & FMR response \\ 
Gallardo \textit{et al.} 2014 [\onlinecite{Gallardo2014}] & ~ & 1D / 2D & ~ & 30 & ~ & 0--10 & ~ & Permalloy & ~ & 0.3 & ~ & TMS Theory & ~ & FMR response\\
Aranda \textit{et al.} 2014 [\onlinecite{Aranda2014}] & ~ & 1D & ~ & 20 & ~ & 0--100 & ~ & Permalloy & ~ & 0.5 & ~ & Simulation & ~ & FMR response\\
Gallardo \textit{et al.} 2018 [\onlinecite{Gallardo2018}] & ~ & 1D & ~ & 10 & ~ & 0--50 & ~ & Permalloy & ~ & 0.3 & ~ & PWM Theory & ~ & FMR response\\
\hline\hline
\end{tabular}
\end{table*}

The etching depths of each SMMC were determined by fitting of the $f(H_0)$ dependence of SW modes standing vertically in the film, also referred to as perpendicular standing spin-wave (PSSW) modes.\cite{Tannenwald1957,Seavey1959} For larger modulation heights $\Delta d > 10$~nm, vibrating sample magnetometry measurements were employed to determine the magnetic volume of an etched reference film before ($M_1$) and after ($M_2$) the ion milling step. This approach estimates the modulation height $\Delta d=d\times\left(1-M_2/M_1\right)$ by the ratio of the measured magnetic volume $M_2/M_1$. 

The following discussion of SW modes in SMMCs is subdivided by the in-plane orientation of the external field with respect to the modulation axis of the SMMC, i.e.\ backward-volume (BV) and Damon-Eshbach (DE) geometry.
%
%

%
%
%

\section{Backward-volume geometry}
\label{BV}

SWs in an SMMC with field orientation parallel to the modulation axis, i.e.\ $\mathbf{k} \| \mathbf{M}$ (BV geometry), are particularly interesting due to the highly inhomogeneous demagnetizing fields $H_{\mathrm{d}}$, which are largest compared to all other in-plane field orientations. Before different transitional SMMCs are addressed, both, the planar film limit as well as the limit of a `full' MC of separate wires are addressed. 

In order to understand the complex mode structure of the SMMCs, micromagnetic simulations were carried out using the MuMax$^3$-code.\cite{Vansteenkiste2014} The FMR simulations were carried out employing both, pulsed\cite{McMichael2005} and continuous-wave\cite{Wagner2015} excitation. Both approaches are described in more detail in Ref.~\onlinecite{Langer2017}. For all simulations, the material parameters presented in Sec.~\ref{Exp} were used and several SMMCs were micromagnetically reconstructed using a fixed film thickness of $d=36$~nm and six modulation heights of $\Delta d=2, 4, 9, 13.5, 18$ and 36~nm with the latter representing the `full' MC. For the realization of an extended MC, periodic boundary conditions were applied to the in-plane axes.

\subsection{The Limits: Thin Film and Full Magnonic Crystal}

\vspace{3mm}
\textbf{Thin Film Limit.}\quad The limit of a thin film with perturbation-like modulations has previously been studied in experiment\cite{Barsukov2011,Koerner2013,Langer2016} and theory using two-magnon scattering perturbation theory.\cite{Landeros2012,Gallardo2014} In such systems, the demagnetizing field $H_{\mathrm{d}}$ acts as a small periodic perturbation introducing a base periodicity $a_0$ crucial for the presence of standing (in-plane) SW modes. While the dispersion remains unaffected, an introduction of periodic perturbations in the real space leads to a discretization of standing SW modes in the $k$-space according to $k=2\pi n/a_0$ with $k$ being the in-plane wave vector and $n=1, 2, 3,...$ the film mode ``quantum number''. As mentioned above, the SW dispersion follows the well known relation:\cite{Kalinikos1986}
%
%
\begin{align}
\left(\frac{\omega}{\gamma}\right)^{\negmedspace2}\negthickspace&=\!\left[ \mu_{0} H_\mathrm{eff} + D k^2 \right]\label{BVdisp}\\\nonumber
&~\times \left[ \mu_{0} H_\mathrm{eff} + \mu_{0} M_{\mathrm{s}} \left( \frac{1-\mathrm{e}^{-kd}}{kd} \right) + D k^2 \right]
\end{align}
\noindent with the FMR frequency $f = \omega/(2\pi)$ and the effective field $H_\mathrm{eff}=H_\mathrm{0}+H_\mathrm{d}$ composed of the external field $H_{0}$ and the internal demagnetizing field $H_{\mathrm{d}}$.

Equation~(\ref{BVdisp}) describes a parabolic shape of the dispersion for dipole-exchange spin waves in BV geometry with minimum frequency at $k \neq 0$. The characteristic shape implies an energy degeneracy, which together with the scattering condition $k=2\pi n/a_0$ enables the two-magnon scattering channel\cite{Landeros2012,Gallardo2014} from the uniform mode to $k\neq 0$ standing SW modes. This is the reason for the occurrence of many high-intensity modes at the same time in contrast to the DE geometry (see Sec.~\ref{DE}).

In Figs.~\ref{FilmLimit}(a) and \ref{FilmLimit}(c), the measured and the simulated frequency-field dependences $f(H_0)$ are depicted for a flat film and in Figs.~\ref{FilmLimit}(b) and \ref{FilmLimit}(d) for an SMMC with 2~nm modulation height. Fig.~\ref{FilmLimit}(d) additionally provides the $f(H_0)$ dependence of the $n=1...4$ modes calculated from Eq.~(\ref{BVdisp}) and plotted as orange solid lines. Clearly, standing SW modes in an SMMC with tiny modulation follow the $f(H_0)$ dependence of propagating modes in a flat film of the same properties quite well. However, deviations from the film limit can be observed in the vicinity of crossing points [see crossing between dashed and solid lines in Fig.~\ref{FilmLimit}(b)] in the $f(H_0)$ dependence, where different standing SW modes couple to each other. 

In analogy to coupled oscillators in mechanics, both modes split up into an acoustical and an optical branch with in-phase and anti-phase oscillation, respectively. The evolving frequency gap between acoustical and optical mode is connected with the emergence of a magnonic band structure reflecting the transition from pure film physics to the physics of MCs.
%
%

\begin{figure}[t]
\includegraphics[width=0.95\columnwidth]{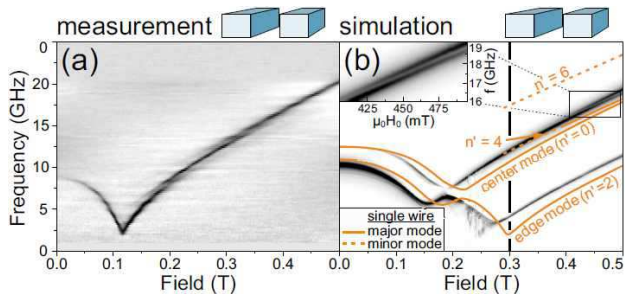}
\caption{\label{WireLimit} (Color online) (a) Measurement and (b) simulation of the frequency-field dependence of an array of wires (one-dimensional MC) with the orange lines indicating the simulated resonance branches of a single wire with the same properties. The edge mode is clearly suppressed in the measurement due to tapered side-walls and side-wall roughness of the measured sample. The vertical line marks the resonance positions at 300~mT shown in Fig.~\ref{wiremodes}.}
\end{figure}

\vspace{3mm}
\textbf{Full MC.}\quad The opposite limit is the periodic array of wires, also known as `full' or one-dimensional MC,\cite{Tacchi2012,Gubbiotti2010} with the $f(H_0)$ dependence shown in Fig.~\ref{WireLimit}. Figure~\ref{WireLimit}(a) is the measurement and \ref{WireLimit}(b) the simulation with orange lines indicating the simulated resonance modes of a single wire. If the stray field contribution of neighboring wires (approximately 30~mT) is considered to shift the modes to lower resonance fields, both the full MC [colorplot in Fig.~\ref{WireLimit}(b)] and the single wire [orange lines in Fig.~\ref{WireLimit}(b)] reveal very similar results for the given dimensions. 

Figure~\ref{wiremodes} contains the corresponding mode profiles of the simulated spin-wave resonances at 300~mT. For the full MC, the mode number is denoted as $n'$ 
and the mode profiles of standing spin-waves are distorted by the strong locally varying internal demagnetizing fields [e.g., the \textit{center} mode ($n'=0$) and the \textit{edge} mode ($n'=2$) displayed in Fig.~\ref{wiremodes}(a) and (b)]. In particular the second mode ($n'=2$) is strongly distorted by the presence of two energetic minima (spin-wave wells\cite{Bayer2003, Bayer2005}) at the edges of the wire created by the strong demagnetizing fields at this location. It is important to note that the amount of \textit{localized modes} in the edge regions is given by the number of SW resonances below the quasi-uniform mode ($n'=0$), which is the first mode with the ability to overcome the maximum of the demagnetizing field in the center and, thus, to extend over the complete structure. It is only possible due to the characteristic parabolic dispersion in the backward-volume geometry that modes of a higher mode number $n'$ bear less energy than the quasi-uniform mode. 

Typical examples for extended modes are found at higher energies, and their mode profiles are illustrated in Figs.~\ref{wiremodes}(c) and \ref{wiremodes}(d). These modes are clearly of a higher order, which is reflected by the number of $n'$ nodes and $n'+1$ peaks inside the magnetic wires and they carry only higher intensities when crossing the center mode to which they can couple [see the inset in Fig.~\ref{WireLimit}(b)]. Note that a symmetric excitation only allows for the measurement of symmetric SW eigenmodes and, thus, the peak number $n'+1$ of SW modes in the `full' MC limit must be odd ($n'=0, 2, 4,...$).\cite{Demokritov2001, Demokritov2003} 
%

\subsection{Transitional SMMCs}

To explain the mode character in hybrid structures with a significant modulation height, two spin-wave ``quantum numbers'' need to be introduced. $l=1, 3, 5,...$ defines the number of peaks of a spin-wave mode inside the thin part of the SMMC and $m=1, 3, 5,...$ reflects the number of peaks within the thick part. In the film limit, the two quantum numbers are linked by the discrete film mode ``quantum number'' $n$ with $2n=m+l$ whereas in the full MC only the $m$ peaks inside the wire region remain connecting $m$ with the full magnonic crystal quantum number $n'$ by $n'=m-1$.
\begin{figure}[t]
\includegraphics[width=1\columnwidth]{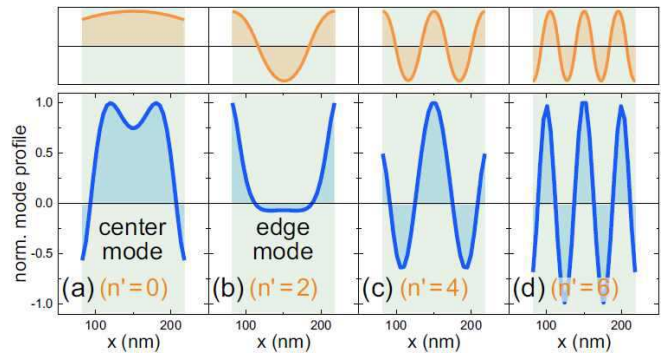}
\caption{\label{wiremodes} (Color online) Simulated (bottom row) and idealized (top row) mode profiles in an array of wires in backward-volume geometry [as shown in Fig.~\ref{WireLimit}(b)] at 300~mT. The modes are labeled according to their mode number $n'$.}
\end{figure}

\begin{figure}[t]
\includegraphics[width=0.95\columnwidth]{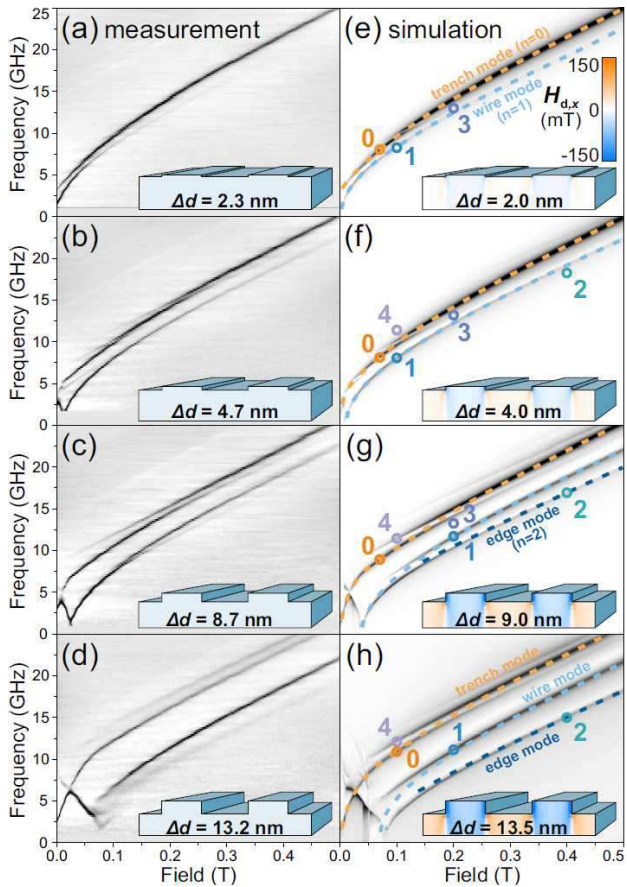}
\caption{\label{multistep} (Color online) (a)--(d) FMR measurements of MCs with different modulation heights as sketched in the insets. (e)--(h) The dynamic response with the internal demagnetizing field in the inset both calculated using pulsed micromagnetic simulations. Colored dots represent the field-frequency position of the mode profiles illustrated in Fig.~\ref{profiles}.}
\end{figure}

Figure~\ref{multistep} shows the $f(H_0)$ plots of several SMMCs with a modulation height $\Delta d$ in the range of 2.0--13.5~nm. In Figs.~\ref{multistep}(a)--(d) and \ref{multistep}(e)--(h), the FMR measurements and the simulations are illustrated, respectively. In the inset, the level of the modulation is provided together with the $x$-component [axes defined according to Fig.~\ref{experiment}(a)] of the simulated internal demagnetizing field $H_{\mathrm{d},x}$ as colorplot. Inside the thick part (also termed `wire') of the MC, the demagnetizing field acts \textit{against} (negative sign) the local magnetization direction as usually. Interestingly, inside the thin part (`trench'), it acts \textit{with} the magnetization and is, thus, a \textit{magnetizing} field. The increasing contrast between the magnetizing- and demagnetizing fields with the modulation height is easily inferred from the insets in Figs.~\ref{multistep}(e)--(h). 

\begin{figure}[!ht]
\includegraphics[width=1\columnwidth]{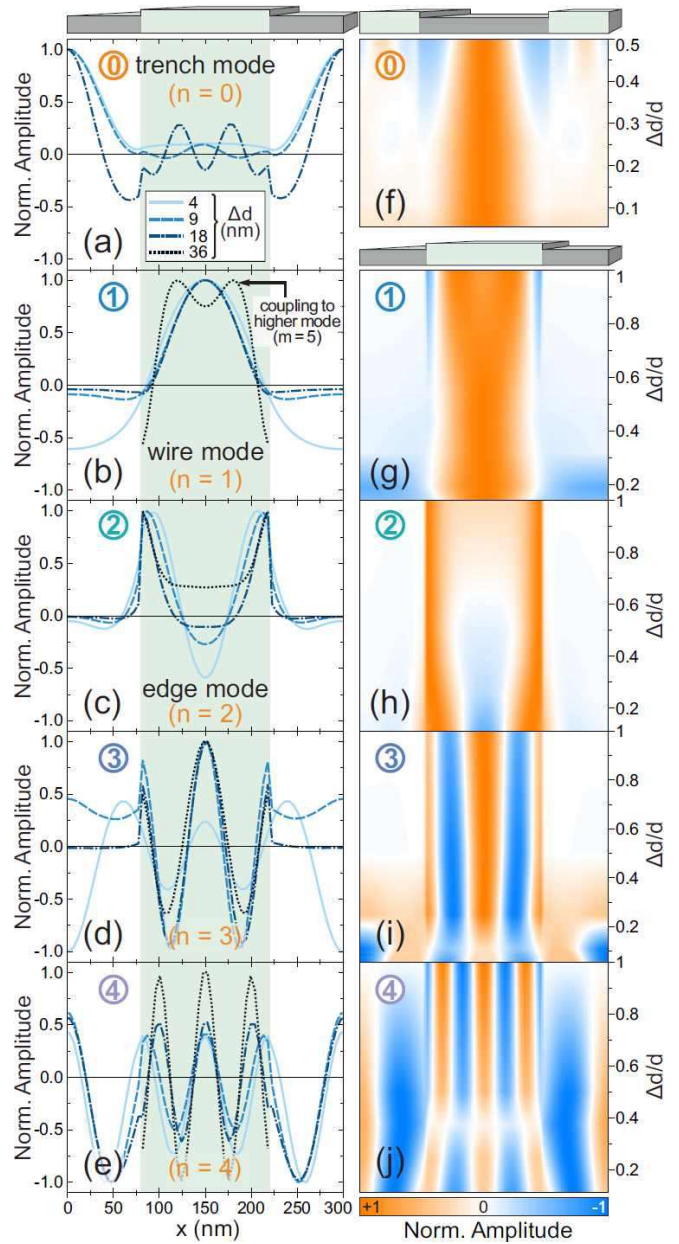}
\caption{\label{profiles} (Color online) (a)--(e) Mode profiles at the marked resonance positions in Fig.~\ref{multistep} revealing the transitional states between the film and the full MC limit. (g)--(j) The $\Delta d$-dependent mode profiles inside the wire region being an important proof for the transition from the $n^{\mathrm{th}}$ film mode to the $m^{\mathrm{th}}$ mode in the full MC limit.}
\end{figure}

The corresponding mode profiles for the marked resonance positions in Fig.~\ref{multistep} are provided in Fig.~\ref{profiles}. Before discussed in detail, the modes are distinguished into three categories: fundamental modes, localized modes and extended modes. The latter two categories are distinguished by whether or not the mode spreads over the complete structure whereas a fundamental mode defines the lowest (center) mode within one of the two different regions, i.e.\ wire or trench.

%

\vspace{3mm}
\textbf{Quasi-Uniform Mode.}\quad A classical uniform mode does no longer exist in an SMMC due to the inhomogeneous internal demagnetizing fields. However, from the mode profiles in Figs.~\ref{profiles}(a) and \ref{profiles}(f) it is evident that the uniform mode evolves to the fundamental mode of the trench region (termed ``trench mode''), i.e.\ a local quasi-uniform excitation of the trench region of the SMMC. In other words, it seems that the quasi-uniform ($n=0$) mode represents the first mode which is able to excite the complete SMMC. Therefore, the mode needs sufficient energy to overcome the magnetizing fields inside the trench region. One consequence is, that the main amplitude of this mode is found inside the trench region. Due to the higher energy necessary to excite the trench, the resonance condition does not match for the wire region anymore. It has been observed that the magnetizing fields inside the trench region increase with the level of surface modulation separating the quasi-uniform mode even stronger from the energetically lower modes which localize only in the wire region. This evolution is reflected by the increasing gap between the trench (orange line) and wire mode (bright blue line) in the $f(H_0)$-dependence illustrated in Figs.~\ref{multistep}(e)--(h). The concentration of the main amplitude of the mode inside the trench region can be seen in the corresponding mode profiles shown in Fig.~\ref{profiles}(a) and (f) at 75--100~mT. Throughout the mode's evolution, it maintains its main amplitude in the center of the trench region. Thus, in principle the trench mode can be understood as an fundamental $l=0$ excitation of the trench region. But as the coupling to higher SW resonances increases with the modulation height, the uniform ($l=0$) characteristics are gradually reduced and are replaced by a modulus due to the strong hybridization with higher BV modes crossing the $f(H_0)$-dependence of the mode. This leads to the effect that the quasi-uniform branch rather appears as an $l=1,3,5,...$ mode changing the peak number $l$ by $\pm 2$ at each interaction point with a higher SW resonance in the $f(H_0)$-dependence. The crossing point in Figs.~\ref{multistep}(g),(h) at approximately $f=15$~GHz and $\mu_{0} H_\mathrm{0}=200$~mT is such an example. Here the 4$^{\mathrm{th}}$ mode crosses the quasi-uniform branch revealing $l=1$ peaks below the gap and $l=3$ peaks above.

It can be concluded that the quasi-uniform mode evolves towards a fundamental trench mode ($l=0$) which gradually splits up into discreet standing BV modes with $l=1,3,5...$ peaks confined inside the trench region. Moreover, the trench mode gradually loses intensity due to the reduction of the film thickness in the trench region until the full MC limit is reached and the mode completely disappears. This transition is also accompanied by an increasingly strong magnetizing field in the trench region, shifting the trench mode to higher field and frequency values supporting the formation of a large gap with respect to 1$^{\mathrm{st}}$ SW mode [see orange dashed line in Figs.~\ref{multistep}(e)--(h)].
\vspace{3mm}
$\mathbf{1^{\mathrm{st}}}$ \textbf{BV Mode.}\quad The bright blue dashed lines in the Figs.~\ref{multistep}(e)--(h) indicate the wire mode. Originating from the 1$^{\mathrm{st}}$ BV mode in the film limit, [see $n=1$ mode in the inset of Fig.~\ref{FilmLimit}(d)], this mode evolves from an extended BV mode with the wavelength $\lambda = a_0$ to a fundamental (center) mode of the wire with one peak ($m=1$) confined in this region. The mode profiles in Figs.~\ref{profiles}(b) and \ref{profiles}(g) demonstrate this transition. Note that the additional peaks visible for $\Delta d = 36.0$~nm are due to a strong coupling to the $n'=4$ mode [also visible in Figs.~\ref{WireLimit}(a) and \ref{wiremodes}(c)]. 

The reason for the mode's confinement in the wire region lies in the energy of the mode. Already in the film limit, the resonance frequency of the mode lies for all field values $>50$~mT below the uniform mode. With insufficient energy for an excitation of the complete structure, the mode localizes in those regions where the energy is internally reduced due to the presence of a demagnetizing field. Thus, the mode is strongly suppressed in regions with \textit{magnetizing} fields (i.e.\ the trench). Even at low modulation heights of only a few nanometers, the demagnetizing fields in the wire region are strong enough to shift the complete mode to higher field values [bright blue dashed line in Figs.~\ref{multistep}(e)--(h)] with the formation of a large frequency and field gap with respect to the quasi-uniform mode [orange dashed line in Figs.~\ref{multistep}(e)--(h)]. As the locally alternating demagnetizing fields increase with the modulation height, the gap between both fundamental modes of the wire and the trench becomes very large reaching nearly 5~GHz at $\Delta d=13.5$~nm.

Eventually, the 1$^{\mathrm{st}}$ BV mode ($n=1$) becomes therefore the wire mode with $m=2n-1=1$ peaks inside the wire. The mode profiles in Figs.~\ref{profiles}(b) and \ref{profiles}(g) support this transition of the mode character consistently up to the limit of the full MC [Fig.~\ref{wiremodes}(b)]. Here the rule $n'=m-1$ applies and the mode forms the 'new' quasi-uniform mode ($n'=0$) of the full MC.
%

\vspace{3mm}
$\mathbf{2^{\mathrm{nd}}}$ \textbf{BV Mode.}\quad 
The transition of the $2^{\mathrm{nd}}$ BV mode is particularly interesting because the mode reveals with its $2n=4$ peaks distributed over a full period $a_0$ a wavenumber which is close to the minimum of the parabolic SW dispersion in backward-volume geometry [Eq.~(\ref{BVdisp})]. This means that the mode does not only have less energy than the quasi-uniform preventing the mode from extending over the full SMMC. The energy is even below the fundamental wire mode (with $m=1$), meaning the $2^{\mathrm{nd}}$ BV mode cannot fully extend over the wire region and is forced to mainly localize at the edges of the wire where the energy is lowest due to the high demagnetizing fields. 

This is the reason why the mode profile of this spin-wave mode is highly distorted and its evolution can be divided into two steps: (i) The localization of $m=2n-1=3$ peaks inside the wire and subsequently, (ii) the gradual suppression of the central peak and the localization at the two wire edges. The reason for (i) is equivalent to the explanation for the localization of the wire mode. In the film limit, the 2$^{\mathrm{nd}}$ SW mode lies even below the 1$^{\mathrm{st}}$ mode (at fields above 100~mT) with the same consequence that the mode cannot extend over the complete structure and instead localizes $m=3$ peaks in the wire region. The reason for (ii) is the formation of the distinct spin-wave wells,\cite{Bayer2005} i.e., minima of $H_{\mathrm{d}}(x)$, coinciding with the outermost peaks of this mode close to the edges of the wire and scaling with the degree of surface modulation. Thus, the mode profiles of the second mode in Fig.~\ref{profiles}(c) and \ref{profiles}(h) are interpreted as a transition from the $n=2$ mode to an $m=2n-1=3$ mode with increasing mode localization at the edges ending up as the edge mode in the full MC [see Fig.~\ref{wiremodes}(a)] with the mode number $n'=m-1=2$.
%
%

\vspace{3mm}
\textbf{Higher BV Modes.}\quad Backward-volume modes of higher order ($n>2$) are also present throughout the transition from a film to a full MC. And as mentioned before, these modes only carry significant intensity when crossing one of the fundamental modes (see Fig.~\ref{multistep}) to which they can couple. Thus, these modes exhibit quite different mode profiles depending on whether they couple to the trench or to the wire mode. In the former case, the modes extend over the complete structure with the largest amplitude in the trench region. In contrast to localized modes, extended modes conserve their total peak number $2n=m+l$ under a gradual increase of the ratio $m/l$ due to the internal field contrast between trench and wire scaling with the surface modulation. In other words, the modes adopt their wavelength to the underlying energy landscape such that the wavelength is reduced where the internal field is high (trench) and vice versa where it is small (wire/edges). This is supported by the mode profiles of the 4$^{\mathrm{th}}$ mode [Fig.~\ref{profiles}(e) and \ref{profiles}(j)] where with increasing modulation height the peaks squeeze up inside the wire region whereas they get wider in the trench region. In the case of higher modes coupling to the wire mode, the modes localize such that they concentrate their amplitude inside the thick part [see profiles of mode 3 and 4 in Figs.~\ref{profiles}(d) and \ref{profiles}(e), respectively]. However, for the transition to the full MC, only the profile inside the wire region is crucial revealing the steady evolution of the mode character to the full MC. This steady process is depicted in Figs.~\ref{profiles}(i) and \ref{profiles}(j) revealing the concentration of $2n-1$ peaks inside the wire.

In brief, the coherent pictures is that the $n^{\mathrm{th}}$ film mode transforms into the $m^{\mathrm{th}}$ mode with $m=2n-1$ during the transition from the film to a highly modulated SMMC. Together with an $l=1$ `silent' (damped) peak inside the etched trench region, this relation reflects the modes' momentum conservation fitting $2n$ peaks inside one period $a_0$ during this process. Arriving in the full MC limit, where only the wire region remains, the simple relation $n'=2(n-1)$ is found as the link to the SW modes in the film limit. Only the $n=0$ mode maintains a local quasi-uniform character inside the trench region and vanishes completely in the full MC.

\section{Damon-Eshbach geometry}
\label{DE}
The DE orientation ($\mathbf{k} \bot \mathbf{M}$) with the field applied parallel to the surface edges offers different insights. Since the static internal demagnetizing fields vanish in this geometry, the transition of spin-wave modes is much more influenced by the dynamic fields and geometrical conditions.

\subsection{The Limits: Thin Film and Full Magnonic Crystal}

\begin{figure}[t]
\includegraphics[width=1\columnwidth]{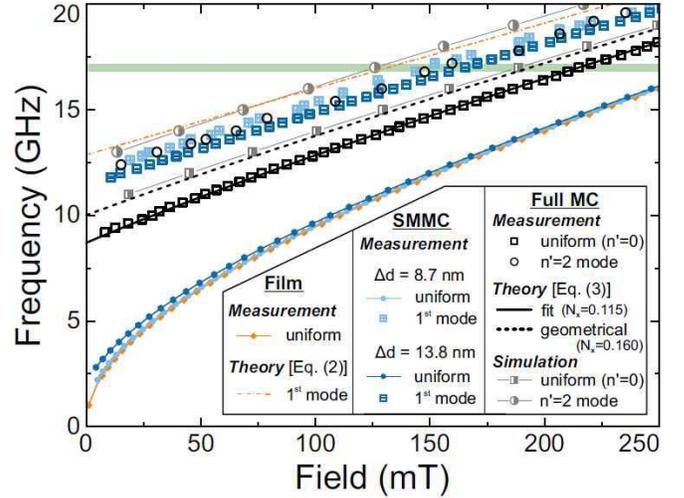}
\caption{\label{multistepDE} (Color online) $f(H_0)$-dependence of SW modes in DE geometry at different surface modulations. The measured quasi-uniform mode (first DE mode) is indicated by full symbols ($\boxplus/\boxminus$ symbols). Black open symbols correspond to the measurement of the full MC. Grey semi-filled symbols represent the simulation of the full MC. Theoretical calculations of the uniform mode in the full MC based on Eq.~(\ref{uniform}) are shown as black solid/dashed lines. The orange dot-dashed line indicates the calculated 1$^{\mathrm{st}}$ SW mode in the film limit [Eq.~(\ref{DEdisp})].}
\end{figure}

\vspace{3mm}
\textbf{Thin Film Limit.}\quad The measurements at lowest modulation heights of $\Delta d=$~2.3~nm, and 4.7~nm did not reveal any significant changes of the FMR spectra compared to the continuous thin film with the typical Kittel-like behavior of the uniform mode. 
Figure~\ref{multistepDE} displays the position of the uniform film mode (orange symbols) together with the calculated position of the 1$^{\mathrm{st}}$ DE mode in the film limit (orange dot-dashed line) according to Eq.~(\ref{DEdisp}). It is noted that the 1$^{\mathrm{st}}$ DE mode carries no intensity in the film limit and a prominent surface-modulation of approximately $d/4$ was necessary for the FMR detection of this mode. 

The appearance of higher non-uniform modes at much higher modulations $\Delta d$ compared to the BV geometry can be explained by the DE SW dispersion\cite{Kalinikos1986,Gallardo2014} reading
\begin{equation}
\begin{split}
\left(\frac{\omega}{\gamma}\right)^{\negmedspace2}\negthickspace=\!&\left[ \mu_{0} H_{\mathrm{eff}} + \mu_{0} M_{\mathrm{s}} \left( 1-\frac{1-\mathrm{e}^{-kd}}{kd} \right) + D k^2 \right]\\
&\times \left[ \mu_{0} H_{\mathrm{eff}} + \mu_{0} M_{\mathrm{s}} \left( \frac{1-\mathrm{e}^{-kd}}{kd} \right) + D k^2 \right]~.
\end{split}
\label{DEdisp}
\end{equation}
In contrast to the BV geometry, there is no energy degeneracy of the uniform mode with higher modes. Thus, no two-magnon scattering channel is present that could easily transfer intensity to higher $k\neq 0$ SW modes.
%

\vspace{3mm}
\textbf{Full MC.}\quad In this limit, the mode profiles are much less distorted compared to the BV geometry due to the vanishing static demagnetizing fields. Thus, the modes can be more easily assigned to the corresponding idealized profile illustrated in the top row of Fig.~\ref{wiremodes}. The main (quasi-uniform) mode with $n'=0$ can be calculated using the demagnetizing factor of the wires\cite{Lindner2004,Belmeguenai2016}
%
%
%
%
\begin{equation}
\begin{split}
\left(\frac{\omega}{\gamma}\right)^{\negmedspace2}\negthickspace=&\left[ \mu_{0} H_{\mathrm{eff}} + N_{x}\mu_{0} M_{\mathrm{s}} \right]\\
&\times \left[ \mu_{0} H_{\mathrm{eff}} - N_x\mu_{0} M_{\mathrm{s}} + \mu_{0} M_{\mathrm{s}} \right]
\end{split}
\label{uniform}
\end{equation}
\noindent with $N_{x}$ being the in-plane demagnetizing factor along the short wire ($x$-) axis [as defined in Fig.~\ref{experiment}(a)] and with the shape-anisotropy $N_x\mu_{0} M_{\mathrm{s}}$. 
Again, higher $k\neq 0$ modes are also present with the wave vector quantized with $k=n'\pi/w$, $n'=2,4,6...$ due to the number of nodes fitting in the wire width $w$.\cite{Belmeguenai2016} It is noticed that the pinning of the mode at the edges of the structure has a strong influence on this estimation and the actual $f(H_0)$ dependence of the mode.

The measurement of the full magnonic crystal (black open symbols in Fig.~\ref{multistepDE}) reveals two modes---the quasi-uniform ($n'=0$) and the 1$^{\mathrm{st}}$ DE mode ($n'=2$). The positions of both modes obtained by micromagnetic simulations are plotted as semi-filled gray squares in Fig.~\ref{multistepDE}). Apart from a systematic reduction of the measured SW resonance frequency compared to the simulation, a qualitative agreement is achieved. It is demonstrated by the $f(H_0)$-dependences calculated with the help of Eq.~(\ref{uniform}) that the frequency of the uniform ($n'=0$) mode is reduced if the effective demagnetizing factor of the wire is smaller. When $N_{x}$ was fitted (solid black in Fig.~\ref{multistepDE}) to the measurement data, a value of $N_x=0.115$ is obtained. $N_{x}$ can also be calculated\cite{Brown1962} employing the ideal geometrical properties of the wires listed in Sec.~\ref{Exp} yielding $N_x=0.160$. As the latter approach (dashed black line in Fig.~\ref{multistepDE}) yields a reasonable agreement of theory and simulation, it can be concluded that imperfect (tapered) edges are likely effectively reducing the demagnetizing factor of the wires leading to the deviation between the simulated and measured $f(H_0)$-dependences.
%
%
\subsection{Transitional SMMCs}
\begin{figure}[t]
\includegraphics[width=1\columnwidth]{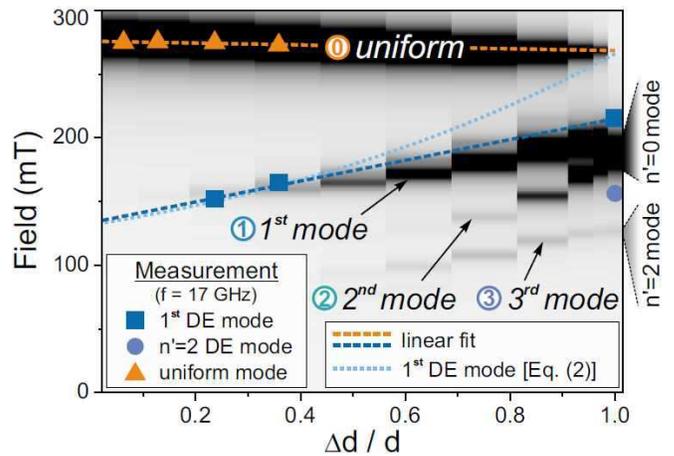}
\caption{\label{transitionDE} (Color online) Micromagnetic simulation of the SW resonances at $f=17$~GHz (green line in Fig.~\ref{multistepDE}) depending on the modulation height $\Delta d$ in DE orientation. The measurement is represented by full symbols. The blue dashed line indicates the position of the 1$^{\mathrm{st}}$ calculated by Eq.~(\ref{DEdisp}) using an effective film thickness $d_{\mathrm{eff}}$.}
\end{figure}

\begin{figure}[t]
\includegraphics[width=0.9\columnwidth]{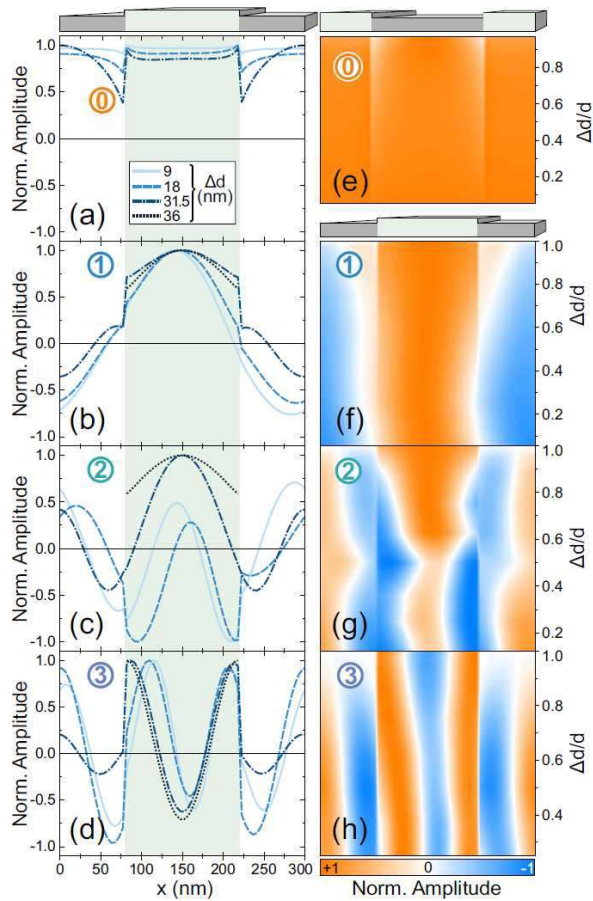}
\caption{\label{ProfilesDE} (Color online) (a)--(d) Simulated spin-wave mode profiles in Damon-Eshbach geometry in the transition from a film to a full MC. (e)--(h) The $\Delta d$-dependent colorplots of the mode profiles integrated over the total film thickness.}
\end{figure}

%
%
%
In Fig.~\ref{multistepDE}, the $f(H_0)$-dependence of the measured $\Delta d$-dependent SW resonances is illustrated for the quasi-uniform mode (full symbols) and the 1$^{\mathrm{st}}$ DE mode (open symbols with and without filling pattern). The measurement is corroborated with the results from theoretical calculations [according to Eqs.~(\ref{DEdisp}) and (\ref{uniform})] of the 1$^{\mathrm{st}}$ mode in the film limit (orange dot-dashed line) and in the full MC limit (black dashed and solid lines). The results from the micromagnetic simulations of the full MC are plotted as gray semi-filled symbols. The green line marks the frequency of 17~GHz at which the field-dependent dynamic response was simulated and compared to the measurement data (Fig.~\ref{transitionDE}) for different modulation heights $\Delta d$. The dynamic response is plotted in gray scale and colored full symbols mark the measured SW resonances. 
The corresponding simulated mode profiles are presented in Fig.~\ref{ProfilesDE} for the quasi-uniform mode and the first three higher modes.


Before the transition process for each of these modes is addressed in detail, it is noted that DE modes in SMMCs show significant dipolar induced non-reciprocal properties.\cite{Lisenkov2015} As a consequence, the $n\neq0$ DE modes reveal a propagating character and their mode profiles are no longer fully symmetric with a tendency to show a non-uniform vertical mode profile in addition. In the following discussion this complexity will not be addressed and the main focus will be on the averaged transversal characteristics of the DE modes.

\vspace{3mm}
\textbf{Quasi-Uniform Mode.}\quad 
In the $f(H_0)$-dependence, the uniform mode maintains its Kittel-like behavior up to significant modulations even above $\Delta d=d/2$, as can be seen from Fig.~\ref{multistep} and Fig.~\ref{transitionDE}, which is a main difference compared to the BV geometry, where the uniform mode evolves rapidly towards a trench mode with a strong up-shift of the resonance frequency due to the presence of magnetizing fields. The preservation of its original uniform character is also reflected in the mode profiles shown in Figs.~\ref{ProfilesDE}(a),(e). Apart from the wire edges, the amplitude of the mode is almost evenly spread over a full period even up to prominent modulation heights $\Delta d>d/2$. Interestingly, at very high modulations close to $\Delta d\rightarrow d$ the mode seems to evolve a stronger amplitude in the center of the trench [see mode profile at $\Delta d=31.5$~nm in Fig.~\ref{ProfilesDE}(a)] , similar to the observation in the BV orientation. In Fig.~\ref{multistepDE}, it can be seen that the quasi-uniform mode maintains its approximate $f(H_0)$-position and loses gradually intensity until it vanishes in the full MC limit. Thus, similar to the BV direction, the main FMR mode ($n'=0$) of the full MC does not originate from the quasi-uniform mode of the SMMC.

\vspace{3mm}
$\mathbf{1^{\mathrm{st}}}$ \textbf{DE Mode.}\quad The measured $f(H_0)$-dependences in Fig.~\ref{multistepDE} indicate instead that the $1^{\mathrm{st}}$ DE mode ($\boxplus/\boxminus$ symbols in Fig.~\ref{multistepDE}) gradually evolves to the uniform mode ($n'=0$) of the full CM (open squares in Fig.~\ref{multistepDE}). Clearly, at the modulations of $\Delta d=8.7$~nm and $\Delta d=13.2$~nm the $1^{\mathrm{st}}$ mode gains intensity and is slightly shifted towards lower frequencies and higher field values closer to the position of the quasi-uniform mode of the full MC. This supposed transition is supported by the micromagnetic simulation (colorplot) in Fig.~\ref{transitionDE}. Here, a steady $\Delta d$-dependent development of the 1$^{\mathrm{st}}$ SMMC mode towards the $n'=0$ full MC mode is identified together with the experimentally obtained data (full squares) at 17~GHz. For small modulations, this behavior can be calculated by Eq.~(\ref{DEdisp}) employing a reduced effective film thickness of $d_{\mathrm{eff}}=a^{-1}_0[wd+(a_0-w)(d-\Delta d)]$. This relation is depicted as a dotted blue line in Fig.~\ref{transitionDE} revealing a reasonable agreement for modulations of $\Delta d\leq d/2$. For larger modulations, this estimation becomes systematically wrong and the simulated mode profiles in Figure \ref{ProfilesDE}(b),(f) start to reveal major deviations from the sinusoidal film-mode character and are dominated by a central peak inside the wire region, which is already characteristic for the wire mode in the full MC. 

Similar to the BV geometry, the full evolution of the $1^{\mathrm{st}}$ DE mode goes via a state of a pronounced peak inside the wire region ($m=1$) and a suppression of the dynamic response in the trench region ($l=0$). In the full MC limit, only this one peak inside the wire remains and the mode is identified as the quasi-uniform ($n'=0$) mode.

\vspace{3mm}
$\mathbf{2^{\mathrm{nd}}}$ \textbf{DE Mode.}\quad The intensity plot of the simulated dynamic response in Fig.~\ref{transitionDE} reveals that the 2$^{\mathrm{nd}}$ DE mode appears at much higher modulation compared to the $1^{\mathrm{st}}$ mode. This might as well be the reason why this mode was not observed in the FMR measurement. Moreover, the simulations indicate an interesting detail, namely the merging of the $1^{\mathrm{st}}$ and $2^{\mathrm{nd}}$ DE mode in the full MC limit. Note that the two modes remain as individual resonances even up to a modulation of $\Delta d/d=0.97$. At these high modulations, the two resonances move gradually towards each other until they fully merge into the quasi-uniform mode in the full MC. 

To understand the evolution of this mode and its merging process, the mode profiles in Figs.~\ref{ProfilesDE}(c),(g) need to be considered. Besides the distortions due to the vertical profile, the mode shows its characteristic four peaks within one period for small modulations [see the profile for $\Delta d=9$~nm in Fig.~\ref{ProfilesDE}(c)]. In the regime of small modulations, two peaks of the mode are located at the two wire edge, which is natural for the $2^{\mathrm{nd}}$ DE mode since its wavelength is exactly twice the periodicity $a_0$. But as the modulation height increases, it comes to the critical point where these two peaks either need to be packed into the trench region or the wire region (as observed for the $2^{\mathrm{nd}}$ BV mode) in order to avoid the edges. In contrast to the BV orientation, there are no static demagnetizing fields present and, therefore, the mode is highly influenced by the geometrical properties. That means, that the width of the wire and the trench are likely to play an important role here. To be more precise, as the width of the wire is slightly smaller than the trench, an evolution towards a mode with $m=1$ and $l=3$ is energetically more favorable than a mode with $m=3$ and $l=1$. This is supported by the mode profile for $\Delta d=31.5$~nm in Fig.~\ref{ProfilesDE}(c) revealing that one peak inside the wire and three peaks in the trench region are established.
\\

\textbf{Higher DE Modes.}\quad gain, there are also higher modes, like the $3^{\mathrm{rd}}$ DE mode, which maintains its characteristic $2n=6$ peaks throughout the transition until the full MC is reached (see Figs.~\ref{ProfilesDE}(d),(h)). In this limit, only the three peaks inside the wire remain, forming the characteristic shape of the $n'=2$ mode with two peaks at the edges and one in the center. 

This seems to be a general transition pattern in the DE geometry. For moderate modulations, the modes maintain their sinusoidal behavior much longer compared to the BV geometry. As the modulation becomes prominent, the modes avoid steady peaks at the edges which are either placed inside the trench or inside the wire, depending on which of both possibilities
\begin{align}
m&=n-1,\quad l=n+1\nonumber \\
\text{or}\quad m&=n+1,\quad l=n-1\nonumber
\end{align}
is closest to the natural wavelength $\lambda=n/a_0$ of the mode. In MCs with almost equal size of wire and trench, this circumstance would be relevant for all even ($n=2,4,6,...$) modes which finally reach the full MC with $n'=m-1$ either as $n'=n-2$ mode or $n'=n$ mode. Modes of an odd number ($n=1,3,5,...$) reveal a node close to the edges such that the number of peaks is exactly cut in halves when the full MC is reached. Using the relation $n'=m-1$, this means that odd modes evolve to the $n'=n-1$ full MC mode.

\section{Conclusion}
\label{Con}
The gradual evolution of SW modes in the transition from a thin film to a full MC has been investigated by FMR measurements and micromagnetic simulations in the BV and DE geometry. 
The uniform mode maintains as a trench mode in BV orientation and as a quasi-uniform mode in the DE geometry until it finally vanishes in the full MC limit for both orientations.

For non-uniform modes, simple transition rules are found. While in the BV orientation, the transition process is governed by the presence of strong magnetizing and demagnetizing fields, which lead to a confinement of $m=2n-1$ peaks inside the wire region, the transition process for DE modes is governed by the preservation of the modes natural wavelength leading to the transition to an $m=n$, $l=n$ mode for odd $n$ and either $m=n-1$, $l=n+1$ or $m=n+1$, $l=n-1$ for even $n$, depending on the geometry of the MC. For all modes, the transition ends in the full MC limit where only the peak number $m$ in the wire region pre-determines the final state with $n'=m-1$. Thus, in the BV geometry the $n^{\mathrm{th}}$ film mode evolves to the $n'=2(n-1)$ mode of the full MC. In the DE geometry, odd film modes with $n=1,3,5,...$ evolve to the $n'=n-1$ full MC mode whereas even modes with $n=2,4,6,...$ can either become the $n'=n-2$ mode or the $n'=n$ mode.

\section{Acknowledgment}

We thank B.\ Scheumann for the film deposition, A.\ Banholzer and C.\ Fowley for the technical support in the patterning process and P.\ C.\ Grubitz and A.\ Jansen for their help with the fitting of the measurement data. The support by the Structural Characterization and Nanofabrication Facilities Rossendorf at IBC and the HZDR Department of Information Services and Computing is gratefully acknowledged. 
This work was supported by the Centers of Excellence with Basal/CONICYT financing (grant no.\ FB0807), CONICYT PAI/ACADEMIA 79140033, FONDECYT 1161403, FONDE-CYT Iniciacion 11170736, CONICYT PCCI (grant no.\ 140051), DAAD PPP ALECHILE (grant no.\ 57136331) and from the Deutsche Forschungsgemeinschaft (grant no.\ {LE2443/5-1}). Funding from the European Union's Horizon 2020 research and innovation programme under the Marie Sk\l{}odowska-Curie grant agreement No.\ 701647 is gratefully acknowledged.
\bibliography{references}

\providecommand{\noopsort}[1]{}\providecommand{\singleletter}[1]{#1}%
\begin{thebibliography}{10}
\let\qq=\"

\bibitem{Vasseur1996}
J.~O. Vasseur, L.~Dobrzynski, B.~Djafari-Rouhani and H.~Puszkarski, Phys. Rev.
  B {\bf 54}, 1043 (1996).

\bibitem{Nikitov2001}
S.~Nikitov, P.~Tailhades and C.~Tsai, J. Magn. Magn. Mater. {\bf 236}, 320
  (2001).

\bibitem{Neusser2009}
S.~Neusser and D.~Grundler, Adv. Mater. {\bf 21}, 2927 (2009).

\bibitem{Kruglyak2010}
V.~V. Kruglyak, S.~O. Demokritov and D.~Grundler, J. Phys. D: Appl. Phys. {\bf
  43}, 264001 (2010).

\bibitem{Gubbiotti2010}
G.~Gubbiotti, S.~Tacchi, M.~Madami, G.~Carlotti, A.~O. Adeyeye and M.~Kostylev,
  J. Phys. D: Appl. Phys. {\bf 43}, 264003 (2010).

\bibitem{Lenk2011}
B.~Lenk, H.~Ulrichs, F.~Garbs and M.~M\"unzenberg, Phys. Rep. {\bf 507}, 107
  (2011).

\bibitem{Krawczyk2014}
M.~Krawczyk and D.~Grundler, J. Phys. Condens. Matter {\bf 26}, 123202 (2014).

\bibitem{Chumak2015}
A.~V. Chumak, V.~I. Vasyuchka, A.~A. Serga and B.~Hillebrands, Nat. Phys. {\bf
  11}, 453 (2015).

\bibitem{Mansurova2017}
M.~Mansurova, J.~von~der Haar, J.~Panke, J.~Walowski, H.~Ulrichs and
  M.~M\"unzenberg, J. Phys.: Condens. Matter {\bf 29}, 214001 (2017).

\bibitem{Ma2012}
F.~S. Ma, H.~S. Lim, V.~L. Zhang, S.~C. Ng and M.~H. Kuok, Nanoscale Res. Lett.
  {\bf 7}, 1 (2012).

\bibitem{Tacchi2012}
S.~Tacchi, G.~Duerr, J.~W. K{\l}os, M.~Madami, S.~Neusser, G.~Gubbiotti,
  G.~Carlotti, M.~Krawczyk and D.~Grundler, Phys. Rev. Lett. {\bf 109}, 137202
  (2012).

\bibitem{Yu2013}
H.~Yu, G.~Duerr, R.~Huber, M.~Bahr, T.~Schwarze, F.~Brandl and D.~Grundler,
  Nat. Commun. {\bf 4}, 2702 (2013).

\bibitem{Saha2013}
S.~Saha, S.~Barman, J.~Ding, A.~O. Adeyeye and A.~Barman, Appl. Phys. Lett.
  {\bf 102}, 242409 (2013).

\bibitem{Mruczkiewicz2013}
M.~Mruczkiewicz, M.~Krawczyk, G.~Gubbiotti, S.~Tacchi, Y.~A. Filimonov, D.~V.
  Kalyabin, I.~V. Lisenkov and S.~A. Nikitov, New J. Phys. {\bf 15}, 113023
  (2013).

\bibitem{Obry2013}
B.~Obry, P.~Pirro, T.~Br\"acher, A.~V. Chumak, J.~Osten, F.~Ciubotaru, A.~A.
  Serga, J.~Fassbender and B.~Hillebrands, Appl. Phys. Lett. {\bf 102}, 202403
  (2013).

\bibitem{Rychly2017}
J.~Rych{\l}y and J.~W. K{\l}os, J. Phys. D: Appl. Phys. {\bf 50}, 164004
  (2017).

\bibitem{Topp2010}
J.~Topp, D.~Heitmann, M.~P. Kostylev and D.~Grundler, Phys. Rev. Lett. {\bf
  104}, 207205 (2010).

\bibitem{Ding2011}
J.~Ding, M.~Kostylev and A.~O. Adeyeye, Phys. Rev. Lett. {\bf 107}, 047205
  (2011).

\bibitem{Duerr2012}
G.~Duerr, K.~Thurner, J.~Topp, R.~Huber and D.~Grundler, Phys. Rev. Lett. {\bf
  108}, 227202 (2012).

\bibitem{Schwarze2013}
T.~Schwarze and D.~Grundler, Appl. Phys. Lett. {\bf 102}, 222412 (2013).

\bibitem{Huber2013}
R.~Huber, M.~Krawczyk, T.~Schwarze, H.~Yu, G.~Duerr, S.~Albert and D.~Grundler,
  Appl. Phys. Lett. {\bf 102}, 012403 (2013).

\bibitem{Montoncello2013}
F.~Montoncello, S.~Tacchi, L.~Giovannini, M.~Madami, G.~Gubbiotti, G.~Carlotti,
  E.~Sirotkin, E.~Ahmad, F.~Y. Ogrin and V.~V. Kruglyak, Appl. Phys. Lett. {\bf
  102}, 202411 (2013).

\bibitem{Barsukov2011}
I.~Barsukov, F.~M. R\"omer, R.~Meckenstock, K.~Lenz, J.~Lindner,
  S.~HemkentoKrax, A.~Banholzer, M.~K\"orner, J.~Grebing, J.~Fassbender and
  M.~Farle, Phys. Rev. B {\bf 84}, 140410 (2011).

\bibitem{Landeros2012}
P.~Landeros and D.~L. Mills, Phys. Rev. B {\bf 85}, 054424 (2012).

\bibitem{Liu2013}
X.~M. Liu, J.~Ding, G.~N. Kakazei and A.~O. Adeyeye, Appl. Phys. Lett. {\bf
  103}, 062401 (2013).

\bibitem{Gallardo2014}
R.~A. Gallardo, A.~Banholzer, K.~Wagner, M.~K\"orner, K.~Lenz, M.~Farle,
  J.~Lindner, J.~Fassbender and P.~Landeros, New J. Phys. {\bf 16}, 023015
  (2014).

\bibitem{Kakazei2014}
G.~N. Kakazei, X.~M. Liu, J.~Ding and A.~O. Adeyeye, Appl. Phys. Lett. {\bf
  104}, 042403 (2014).

\bibitem{Langer2016}
M.~Langer, K.~Wagner, T.~Sebastian, R.~H\"ubner, J.~Grenzer, Y.~Wang,
  T.~Kubota, T.~Schneider, S.~Stienen, K.~Lenz, H.~Schultheiss, J.~Lindner,
  K.~Takanashi, R.~E. Arias and J.~Fassbender, Appl. Phys. Lett. {\bf 108},
  102402 (2016).

\bibitem{Khanal2017}
S.~Khanal, P.~Sherpa and L.~Spinu, IEEE Trans. Magn. {\bf 53}, 1 (2017).

\bibitem{Heimbach2018}
F.~Heimbach, T.~Stückler, H.~Yu and W.~Zhao, J. Magn. Magn. Mater. {\bf 450},
  29  (2018).

\bibitem{Cuykendall1987}
R.~Cuykendall and D.~R. Andersen, Opt. Lett. {\bf 12}, 542 (1987).

\bibitem{Schneider2008}
T.~Schneider, A.~A. Serga, B.~Leven, B.~Hillebrands, R.~L. Stamps and M.~P.
  Kostylev, Appl. Phys. Lett. {\bf 92}, 022505 (2008).

\bibitem{Lee2008}
K.-S. Lee and S.-K. Kim, J. Phys. D: Appl. Phys. {\bf 104}, 053909 (2008).

\bibitem{Khitun2010}
A.~Khitun, M.~Bao and K.~L. Wang, J. Phys. D: Appl. Phys. {\bf 43}, 264005
  (2010).

\bibitem{Khitun2011}
A.~Khitun and K.~L. Wang, J. Appl. Phys. {\bf 110}, 034306 (2011).

\bibitem{Berut2012}
A.~Berut, A.~Arakelyan, A.~Petrosyan, S.~Ciliberto, R.~Dillenschneider and
  E.~Lutz, Nature {\bf 483}, 187 (2012).

\bibitem{Khitun2012}
A.~Khitun, J. Appl. Phys. {\bf 111}, 054307 (2012).

\bibitem{Sato2013}
N.~Sato, K.~Sekiguchi and Y.~Nozaki, Appl. Phys. Express {\bf 6}, 063001
  (2013).

\bibitem{Jamali2013}
M.~Jamali, J.~H. Kwon, S.-M. Seo, K.-J. Lee and H.~Yang, Sci. Rep. {\bf 3},
  3160 (2013).

\bibitem{Klingler2014}
S.~Klingler, P.~Pirro, T.~Brächer, B.~Leven, B.~Hillebrands and A.~V. Chumak,
  Appl. Phys. Lett. {\bf 105}, 152410 (2014).

\bibitem{Kostylev2008}
M.~Kostylev, P.~Schrader, R.~L. Stamps, G.~Gubbiotti, G.~Carlotti, A.~O.
  Adeyeye, S.~Goolaup and N.~Singh, Appl. Phys. Lett. {\bf 92}, 132504 (2008).

\bibitem{Lee2009}
K.-S. Lee, D.-S. Han and S.-K. Kim, Phys. Rev. Lett. {\bf 102}, 127202 (2009).

\bibitem{Wang2009}
Z.~K. Wang, V.~L. Zhang, H.~S. Lim, S.~C. Ng, M.~H. Kuok, S.~Jain and A.~O.
  Adeyeye, Appl. Phys. Lett. {\bf 94}, 083112 (2009).

\bibitem{Wang2010}
Z.~K. Wang, V.~L. Zhang, H.~S. Lim, S.~C. Ng, M.~H. Kuok, S.~Jain and A.~O.
  Adeyeye, ACS Nano {\bf 4}, 643 (2010).

\bibitem{Di2013}
K.~Di, H.~S. Lim, V.~L. Zhang, M.~H. Kuok, S.~C. Ng, M.~G. Cottam and H.~T.
  Nguyen, Phys. Rev. Lett. {\bf 111}, 149701 (2013).

\bibitem{Krawczyk2013}
M.~Krawczyk, S.~Mamica, M.~Mruczkiewicz, J.~W. K{\l}os, S.~Tacchi, M.~Madami,
  G.~Gubbiotti, G.~Duerr and D.~Grundler, J. Phys. D: Appl. Phys. {\bf 46},
  495003 (2013).

\bibitem{Kumar2014}
D.~Kumar, J.~W. K{\l}os, M.~Krawczyk and A.~Barman, J. Appl. Phys. {\bf 115},
  043917 (2014).

\bibitem{Inoue2011}
M.~Inoue, A.~Baryshev, H.~Takagi, P.~B. Lim, K.~Hatafuku, J.~Noda and K.~Togo,
  Appl. Phys. Lett. {\bf 98}, 132511 (2011).

\bibitem{Rana2018}
B.~Rana and Y.~Otani, Phys. Rev. Appl. {\bf 9}, 014033 (2018).

\bibitem{Kim2009}
S.-K. Kim, K.-S. Lee and D.-S. Han, Appl. Phys. Lett. {\bf 95}, 082507 (2009).

\bibitem{Vogt2014}
K.~Vogt, F.~Fradin, J.~Pearson, T.~Sebastian, S.~Bader, B.~Hillebrands,
  A.~Hoffmann and H.~Schultheiss, Nat. Commun. {\bf 5}, 3727 (2014).

\bibitem{Balinskiy2018}
M.~Balinskiy, H.~Chiang and A.~Khitun, AIP Adv. {\bf 8}, 056628 (2018).

\bibitem{Chumak2014}
A.~V. Chumak, A.~A. Serga and B.~Hillebrands, Nat. Commun. {\bf 5}, 4700
  (2014).

\bibitem{Bauer2012}
G.~E.~W. Bauer, E.~Saitoh and B.~J. van Wees, Nat. Mater. {\bf 11}, 391 (2012).

\bibitem{Urazhdin2014a}
S.~Urazhdin, V.~E. Demidov, H.~Ulrichs, T.~Kendziorczyk, T.~Kuhn, J.~Leuthold,
  G.~Wilde and S.~O. Demokritov, Nat. Nanotech. {\bf 9}, 509 (2014).

\bibitem{Liu2018}
C.~Liu, J.~Chen, T.~Liu, F.~Heimbach, H.~Yu, Y.~Xiao, J.~Hu, M.~Liu, H.~Chang,
  T.~Stueckler, S.~Tu, Y.~Zhang, Y.~Zhang, P.~Gao, Z.~Liao, D.~Yu, K.~Xia,
  N.~Lei, W.~Zhao and M.~Wu, Nat. Commun. {\bf 9}, 738 (2018).

\bibitem{Vogel2015}
M.~Vogel, A.~V. Chumak, E.~H. Waller, T.~Langner, V.~I. Vasyuchka,
  B.~Hillebrands and G.~von Freymann, Nat. Phys. {\bf 11}, 487 (2015).

\bibitem{Chumak2009-JPD}
A.~V. Chumak, T.~Neumann, A.~A. Serga, B.~Hillebrands and M.~P. Kostylev, J.
  Phys. D: Appl. Phys. {\bf 42}, 205005 (2009).

\bibitem{Chumak2010}
A.~V. Chumak, V.~S. Tiberkevich, A.~D. Karenowska, A.~A. Serga, J.~F. Gregg,
  A.~N. Slavin and B.~Hillebrands, Nat. Commun. {\bf 1}, 141 (2010).

\bibitem{Bai2011}
L.~Bai, M.~Kohda and J.~Nitta, Appl. Phys. Lett. {\bf 98}, 172508 (2011).

\bibitem{Wang2017}
Q.~Wang, A.~V. Chumak, L.~Jin, H.~Zhang, B.~Hillebrands and Z.~Zhong, Phys.
  Rev. B {\bf 95}, 134433 (2017).

\bibitem{Chumak2008}
A.~V. Chumak, A.~A. Serga, B.~Hillebrands and M.~P. Kostylev, Appl. Phys. Lett.
  {\bf 93}, 022508 (2008).

\bibitem{Chumak2009-JAP}
A.~V. Chumak, A.~A. Serga, S.~Wolff, B.~Hillebrands and M.~P. Kostylev, J.
  Appl. Phys. {\bf 105}, 083906 (2009).

\bibitem{Chumak2009-APL}
A.~V. Chumak, A.~A. Serga, S.~Wolff, B.~Hillebrands and M.~P. Kostylev, Appl.
  Phys. Lett. {\bf 94}, 172511 (2009).

\bibitem{Aranda2014}
G.~R. Aranda, G.~N. Kakazei, J.~Gonz\'alez and K.~Y. Guslienko, J. Appl. Phys.
  {\bf 116}, 093908 (2014).

\bibitem{Gallardo2018}
R.~A. Gallardo, T.~Schneider, A.~Rold\'an-Molina, M.~Langer, A.~S.
  N{\'{u}}{\~{n}}ez, K.~Lenz, J.~Lindner and P.~Landeros, Phys. Rev. B {\bf
  97}, 174404 (2018).

\bibitem{Demokritov2001}
S.~Demokritov, B.~Hillebrands and A.~Slavin, Phys. Rep. {\bf 348}, 441  (2001).

\bibitem{Roussigne2001}
Y.~Roussign\'e, S.~M. Ch\'erif, C.~Dugautier and P.~Moch, Phys. Rev. B {\bf
  63}, 134429 (2001).

\bibitem{Gubbiotti2007}
G.~Gubbiotti, S.~Tacchi, G.~Carlotti, N.~Singh, S.~Goolaup, A.~O. Adeyeye and
  M.~Kostylev, Appl. Phys. Lett. {\bf 90}, 092503 (2007).

\bibitem{Tacchi2010}
S.~Tacchi, M.~Madami, G.~Gubbiotti, G.~Carlotti, S.~Goolaup, A.~O. Adeyeye,
  N.~Singh and M.~P. Kostylev, Phys. Rev. B {\bf 82}, 184408 (2010).

\bibitem{Saha2015}
S.~Saha, S.~Barman, Y.~Otani and A.~Barman, Nanoscale {\bf 7}, 18312 (2015).

\bibitem{Belmeguenai2016}
M.~Belmeguenai, M.~Gabor, F.~Zighem, D.~Berling, Y.~Roussign\'e, T.~P. Jr.,
  S.~Ch\'erif, C.~Tiusan, O.~Brinza and P.~Moch, J. Magn. Magn. Mater. {\bf
  399}, 199  (2016).

\bibitem{Langer2017}
M.~Langer, F.~R\"oder, R.~A. Gallardo, T.~Schneider, S.~Stienen, C.~Gatel,
  R.~H\"ubner, L.~Bischoff, K.~Lenz, J.~Lindner, P.~Landeros and J.~Fassbender,
  Phys. Rev. B {\bf 95}, 184405 (2017).

\bibitem{Tannenwald1957}
P.~E. Tannenwald and M.~H. Seavey, Phys. Rev. {\bf 105}, 377 (1957).

\bibitem{Seavey1959}
M.~H. Seavey and P.~E. Tannenwald, J. Appl. Phys. {\bf 30}, S227 (1959).

\bibitem{Vansteenkiste2014}
A.~Vansteenkiste, J.~Leliaert, M.~Dvornik, M.~Helsen, F.~Garcia-Sanchez and
  B.~Van~Waeyenberge, AIP Adv. {\bf 4}, 107133 (2014).

\bibitem{McMichael2005}
R.~D. McMichael and M.~D. Stiles, J. Appl. Phys. {\bf 97} (2005).

\bibitem{Wagner2015}
K.~{Wagner}, S.~{Stienen} and M.~{Farle}, ArXiv  (2015).

\bibitem{Koerner2013}
M.~K\"orner, K.~Lenz, R.~A. Gallardo, M.~Fritzsche, A.~M\"ucklich, S.~Facsko,
  J.~Lindner, P.~Landeros and J.~Fassbender, Phys. Rev. B {\bf 88}, 054405
  (2013).

\bibitem{Kalinikos1986}
B.~A. Kalinikos and A.~N. Slavin, J. Phys. C {\bf 19}, 7013 (1986).

\bibitem{Bayer2003}
C.~Bayer, S.~O. Demokritov, B.~Hillebrands and A.~N. Slavin, Appl. Phys. Lett.
  {\bf 82}, 607 (2003).

\bibitem{Bayer2005}
C.~Bayer, J.~Jorzick, B.~Hillebrands, S.~O. Demokritov, R.~Kouba, R.~Bozinoski,
  A.~N. Slavin, K.~Y. Guslienko, D.~V. Berkov, N.~L. Gorn and M.~P. Kostylev,
  Phys. Rev. B {\bf 72}, 064427 (2005).

\bibitem{Demokritov2003}
S.~O. Demokritov, J. Phys.: Condens. Matter {\bf 15}, S2575 (2003).

\bibitem{Lindner2004}
J.~Lindner, T.~Toli\'nski, K.~Lenz, E.~Kosubek, H.~Wende, K.~Baberschke,
  A.~Ney, T.~Hesjedal, C.~Pampuch, R.~Koch, L.~D\"aweritz and K.~Ploog, J.
  Magn. Magn. Mater. {\bf 277}, 159  (2004).

\bibitem{Brown1962}
J.~W.~F. Brown, {\em ,,Magnetostatic Principles in Ferromagnetism,
  Appendix``\/}, edited by E. P. Wohlfarth (North-Holland, Amsterdam) (1962).

\bibitem{Lisenkov2015}
I.~Lisenkov, D.~Kalyabin, S.~Osokin, J.~Klos, M.~Krawczyk and S.~Nikitov, J.
  Magn. Magn. Mater. {\bf 378}, 313  (2015).

\end{thebibliography}
\bibliographystyle{kl_ms_num}
\end{document}